\documentclass[copyright,creativecommons]{eptcs}
\usepackage{breakurl}             
\usepackage{reo-tikz}
\usepackage{amsfonts}
\usepackage{amssymb,amsmath}
\usepackage{mathrsfs}
\usepackage{latexsym}
\usepackage{algorithmic}
\usepackage{algorithm}
 \numberwithin{algorithm}{subsection}
\usepackage{array}
\usepackage{eucal}

\newcommand{\XStransition}[3]{#1 \, \stackrel{#2}{\longrightarrow}\, #3}

\newcommand{\at}{\mathbf{At}}
\newcommand{\Sync}{{\sf Sync}}
\newcommand{\LossySync}{{\sf LossySync}}
\newcommand{\SyncDrain}{{\sf SyncDrain}}
\newcommand{\FIFO}{{\sf FIFO1}}

\newtheorem{definition}{Definition }
\newtheorem{lemma}{Lemma }
\newtheorem{example}{Example }

\title{A Compositional Semantics for Stochastic Reo Connectors}
\author{
Young-Joo Moon
\qquad
Alexandra Silva
\qquad
Christian Krause
\qquad
Farhad Arbab
\institute{Centrum Wiskunde \& Informatica (CWI), Amsterdam, The Netherlands}
\email{\{yjm,ams,c.krause,farhad\}@cwi.nl}
}
\def\titlerunning{A Compositional Semantics for Stochastic Reo Connectors}
\def\authorrunning{Y.J. Moon, A. Silva, C. Krause, F. Arbab}
\begin{document}
\maketitle

\begin{abstract}
In this paper we present a compositional semantics for the channel-based
coordination language Reo which enables the analysis of quality of service (QoS)
properties of service compositions. For this purpose, we annotate Reo channels
with stochastic delay rates and explicitly model data-arrival rates at the
boundary of a connector, to capture its interaction with the services that comprise its environment.
We propose Stochastic Reo automata as an extension of Reo
automata, in order to compositionally derive a QoS-aware semantics for Reo. We further present a translation of Stochastic Reo automata to
Continuous-Time Markov Chains (CTMCs). This translation enables us to use
third-party CTMC verification tools to do an end-to-end performance analysis of
service compositions.
\end{abstract}

\section{Introduction}
In service-oriented computing (SOC), complex
distributed applications are built by composing existing -- often
third-party -- services using additional coordination mechanisms, such as
workflow engines, component connectors, or tailor-made glue code.
Due to the high degree of heterogeneity and the fact that the owner of the
application is not necessarily the owner of its building blocks,
issues involving quality of service (QoS) properties become increasingly
entangled.
Even if the QoS properties of every individual service and connector are
known, it is far from trivial to determine and reason about
the end-to-end QoS of a composed system in its application context. Yet, the end-to-end QoS of a composed service is often as important as its functional properties in determining its viability in its market.

Reo \cite{Arbab04}, a channel-based coordination language, supports the composition of services, and typically, its semantics is given by Constraint Automata (CA) \cite{BSAR06}. However, CA do not account for the QoS properties and cannot capture the context-dependency \cite{BSAR06} of Reo connectors. To capture context-dependency, Reo automata were introduced in \cite{BCS09}, but they still do not account for the QoS properties. Quantitative Intentional Automata (QIA) were proposed in \cite{QIA09} to account for the end-to-end QoS properties of a Reo connector, but no formal results are readily available on their compositionality.

As our contribution, we present \emph{Stochastic Reo automata} to overcome the shortcomings of CA and QIA, mentioned above: a compositional semantic model for reasoning about the end-to-end QoS properties, as well as handling the context-dependency of Reo connectors. We show that the compositionality results of Reo automata extend to Stochastic Reo automata. We present a translation of Stochastic Reo automata to Continuous-Time Markov Chains~(CTMCs). This allows the use of third-party tools for stochastic analysis. Therefore, this paper shows a compositional approach for constructing Markov Chain~(MC) models of complex composite systems, using Stochastic Reo automata as an intermediate model. In other words, Stochastic Reo automata provides a  compositional framework wherein the corresponding CTMC model of a connector can be derived. This approach, thus, constitutes a compositional framework for modeling and analysis of the QoS properties of complex systems, where our translation derives a CTMC model for complex systems for subsequent analysis by other tools.

\section{Overview of Reo}
\label{sec:reo-overview}

Reo\nocite{BSAR06,CCA05} is a channel-based
coordination model wherein so-called \emph{connectors} are used to coordinate,
i.e., control the communication among, components or services exogenously (from
outside of those components and services). In Reo, complex connectors are
compositionally built out of primitive channels. Channels are atomic
connectors with exactly two ends, which can be either \emph{source} or
\emph{sink} ends. Source ends accept data into, and sink ends dispense data out
of their respective channels. Reo allows channels to be undirected, i.e., to have respectively two source or two sink ends.
\vspace{-2mm}
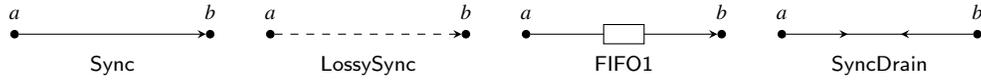
\begin{figure}[h]
\begin{center}
\begin{tikzpicture}[font=\scriptsize]
\node[point] (A) [label=above:$a$] {};
\node[point] (B) [right of = A,xshift=1cm, label=above:$b$] {};
\draw[sync] (A) to node[below,yshift=-1.5mm] {\Sync} (B);

\node[point] (C) [right of = B,xshift=-0.8cm, label=above:$a$] {};
\node[point] (D) [right of = C, xshift=1cm, label=above:$b$] {};
\draw[lossysync] (C) to node[below,yshift=-1.5mm] {\LossySync} (D);

\node[point] (E) [right of = D,xshift=-0.8cm, label=above:$a$] {};
\node[point] (F) [right of = E,xshift=1cm, label=above:$b$] {};
\draw[fifo] (E) to node[below,yshift=-1.5mm] {\FIFO} (F);

\node[point] (G) [right of = F,xshift=-0.8cm, label=above:$a$] {};
\node[point] (H) [right of = G,xshift=1cm, label=above:$b$] {};
\draw[syncdrain] (G) to node[below,yshift=-1.5mm] {\SyncDrain} (H);

\end{tikzpicture}
\end{center}
\vspace{-3mm}
\caption{Some basic Reo channels}
\label{fig:basic_channels}
\end{figure}

Figure \ref{fig:basic_channels} shows the graphical representations
of some basic channel types.  The {\Sync} channel is a directed, unbuffered
channel that synchronously reads data items from its source end and writes
them to its sink end. The {\LossySync} channel behaves similarly, except that it
does not block if the party at the sink end is not ready to receive data.
Instead, it just loses the data item. {\FIFO} is an asynchronous channel with a buffer of size one.
The {\SyncDrain} channel differs from the other channels in that it has two
source ends (and no sink end). If there is data available at both ends, this channel
consumes (and loses) both data items synchronously.

Channels can be joined together using nodes. A node can have one of three types: source, sink or mixed node, depending on whether all ends that coincide on the node
are source ends, sink ends or a combination of both. Source and sink nodes,
called \emph{boundary nodes}, form the boundary of a connector, allowing
interaction with its environment.
%Unlike channels, which are user-defined, the semantics of nodes in Reo is
%fixed.
Source nodes act as synchronous replicators, and sink nodes as mergers.
A mixed node combines both behaviors by atomically consuming a data item
from one sink end and replicating it to all of its source ends.

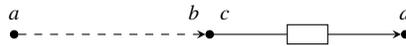
\begin{figure}[h]
\begin{center}
\begin{tikzpicture}[font=\scriptsize]

\node[point] (A) [label=above:$a$] {};
\node[point] (B) [right of=A,xshift=1cm, label=95:$b$, label=85:$c$] {};
\node[point] (C) [right of=B,xshift=1cm, label=above:$d$] {};

\draw[lossysync] (A) to (B);
\draw[fifo]      (B) to (C);

\end{tikzpicture}
\end{center}
\vspace{-3mm}
\caption{Example connector: \sf{LossyFIFO1}}
\label{fig:lossyfifo}
\end{figure}

An example connector is depicted in Figure~\ref{fig:lossyfifo}. It reads a data
item from $a$, buffers it in a {\FIFO} and writes it to $d$. The connector
loses data items from $a$ if and only if the {\FIFO} buffer is already full. This construct
is therefore called (overflow) {\sf LossyFIFO1}.
\vspace{-2mm}
\vspace{-2mm}
\subsection{Semantics: Reo automata}
In this section, we recall Reo automata, an
automata model that provides a compositional operational semantics for
Reo connectors. Intuitively, a Reo automaton is a non-deterministic automaton whose
transitions have labels of the form $g|f$, where $g$
is a {\em guard} (boolean condition) and $f$ a set of nodes that
fire synchronously. A transition can be taken only when its guard $g$ is true.

We recall some facts about Boolean algebras. Let $\Sigma = \{\sigma_1,\ldots,
\sigma_k\}$ be a set of symbols that denote names of connector ports, $\overline{\sigma}$ be the negation of $\sigma$, and ${\cal B}_\Sigma$ be
the free Boolean algebra generated by the following grammar:\vspace{-2mm}
\begin{center}
$g\; ::= \; \sigma\in\Sigma \mid \top \mid  \bot \mid g\vee g \mid g\land g \mid \overline g$
\end{center}

\noindent
We refer to the elements of the above grammar as {\em
guards} and in its representation we frequently omit $\land$ and write
$g_1g_2$ instead of
$g_1\land g_2$. Given two guards $g_1,g_2\in {\cal B}_\Sigma$, we
define
a (natural)
order $\leq$ as
$
g_1\leq g_2 \Longleftrightarrow g_1\land g_2 = g_1
$.
The intended interpretation of $\leq$ is logical implication: $g_1$
implies $g_2$.  An {\em atom} of $\mathcal B_\Sigma$
is a guard  $a_1\ldots a_k$ such that $a_i\in \Sigma \cup \overline{\Sigma}$
with $\overline{\Sigma}=\{\overline{\sigma}_i \mid \sigma_i \in \Sigma\}$,
$1\leq i \leq k$. We can think of an atom as a
truth assignment.  We denote atoms by Greek letters $\alpha,
\beta, \ldots$ and the set of all atoms of $\mathcal B_\Sigma$ by
$\at_\Sigma$.  Given $S\subseteq \Sigma$, we define $\widehat S\in
{\cal B}_\Sigma$ as the
conjunction of all elements of $S$. For instance, for $S = \{a, b,c\}$
we have $\widehat S = abc$.

\begin{definition}%[Reo automaton]
\label{def:reoautomata} \cite{BCS09}
A {\em Reo automaton} is a triple $(\Sigma, Q, \delta)$ where $\Sigma$ is the
set of nodes, $Q$ is the set of states, $\delta \subseteq
Q\times {\cal B}_\Sigma \times 2^\Sigma \times Q$ is the
transition relation such that for each $\XStransition{q}{g|f}{q'}\in\delta$:\\
\begin{tabular}{@{}ll}
~(i) $g \leq\widehat f $ & \qquad (reactivity)\\
(ii) $\forall{g\leq g' \leq \widehat f
}\cdot \forall{\alpha\leq
g'} \cdot \exists{\XStransition{q}{g''|f}{q'}\in\delta}\cdot\ \alpha
\leq g''$ & \qquad (uniformity)
\end{tabular}
\end{definition}
% \vspace{-2mm}
\begin{figure}[t]
\small
\begin{center}
\begin{tabular}{|c|c|c|c|}
\hline
\begin{tikzpicture}[shorten   >=1pt, auto]
         \tikzstyle{every state}=[inner sep=0pt,minimum size=6mm];
                      \node[state] (q)     {$q$};
                   \path[->]
                   (q) edge [out=130,in=50,loop] node {$ab|ab$} ()
                                     ;
        \end{tikzpicture}
&
\begin{tikzpicture}[shorten   >=1pt, auto]
         \tikzstyle{every state}=[inner sep=0pt,minimum size=6mm];
                      \node[state] (q)     {$q$};
                   \path[->]
                   (q) edge [out=130,in=50,loop] node
{$\begin{array}{c}ab|ab\\a\overline
b|a\end{array}$} ()
                                     ;
        \end{tikzpicture}

&
{
\begin{tikzpicture}[shorten   >=1pt, auto]
         \tikzstyle{every state}=[inner sep=0pt,minimum size=6mm];
                      \node[state] (q)     {$q$};
                   \path[->]
                   (q) edge [out=130,in=50,loop] node {$ab|ab$} ()
                                     ;
        \end{tikzpicture}
}
&
{
\begin{tikzpicture}[node distance=2cm, shorten   >=1pt, auto]
         \tikzstyle{every state}=[inner sep=0pt,minimum size=6mm];
          \node[state] (e)           {$e$ };
           \node[state] (f) [right of=e]    {$f$ };

           \path[->]
                   (e) edge[bend left] node[above,sloped]  {$a|a$} (f)
                   (f) edge[bend left] node[below,sloped]  {$b|b$} (e)
                     ;
        \end{tikzpicture}}
\\
\hline
${\sf Sync}$ %--
&
${\sf LossySync}$%--
&
${\sf SyncDrain}$ %--
&${\FIFO}$ %--
\\
\hline
\end{tabular}
\end{center}
\caption{Automata for basic Reo
channels}\label{basic_channels}
\end{figure}
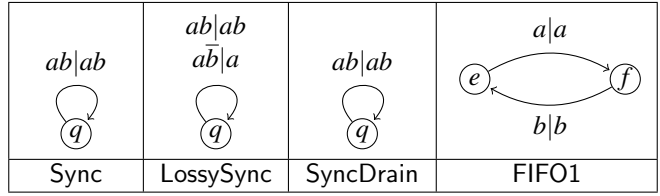

In Reo automata, for simplicity we abstract data constraints \cite{BSAR06} and assume they are \emph{true}. We use arrows $\XStransition{q}{g|f}{q'}$ for $\langle q,g,f,q' \rangle \in \delta$.
If there is more than one transition from
state $q$ to $q'$ we often just draw one arrow and separate their
labels by commas. In Figure~\ref{basic_channels} we depict the Reo
automata for the basic channel types listed in
Figure~\ref{fig:basic_channels}.

 Intuitively,
every
transition
$\XStransition{q}{g|f}{q'}$
in an automaton corresponding to a Reo connector represents that,
if the connector is in state $q$ and the boundary requests present at
the moment,
encoded by an atom $\alpha$, are such that $\alpha \leq g$, then the
nodes $f$ fires
and the connector evolves to state $q'$. Each transition labeled
by $g|f$
satisfies two criteria: (i) \emph{reactivity}---data flows only on nodes
where
a request is pending, capturing Reo's interaction model;
and (ii) \emph{uniformity}---which captures two properties, firstly,
that the request set corresponding precisely to the firing set is
sufficient to cause firing, and secondly,
that removing additional unfired requests from a transition
will not affect the (firing) behavior of the connector \cite{BCS09}.

\subsubsection{Composing Reo connectors}

We now model at the automata level the composition of Reo connectors.
We define two operations: product, which puts two connectors in
parallel, and synchronization, which models the plugging of two nodes. Thus, the product and synchronization operations can be used to obtain the
automaton of a Reo connector by composing the automata of its primitive
connectors. Later in this section we formally show the compositionality of the
operations.

We first define the product operation for Reo automata.
This definition differs from the classical definition of (synchronous) product for
automata: our automata have disjoint alphabets and they can either
take steps together or independently. In the latter case the composite transition in the product automaton
explicitly encodes that one of the two automata cannot perform a step
in the current state, using the following notion:
\begin{definition}\cite{BCS09}
Given a Reo automaton  $\mathcal{A}=(\Sigma, Q, \delta)$ and $q\in
Q$ we define
\begin{center}
$q^\sharp = \neg \bigvee \{~g \mid
\XStransition{q}{g|f}{q'}\in\delta~\}.$
\end{center}
\end{definition}

\noindent
This captures precisely that $\mathcal{A}$ cannot fire in state $q$.

\begin{definition}%[Product]
\cite{BCS09} Given two Reo automata ${\cal A}_1=(\Sigma_1, Q_1, \delta_1)$ and
${\cal
A}_2=(\Sigma_2, Q_2, \delta_2)$ such that $\Sigma_1\cap \Sigma_2
=\emptyset$, we define
the {\em product} of ${\cal A}_1$ and ${\cal A}_2$ as ${\cal
A}_1\times{\cal
A}_2 = (\Sigma_1\cup\Sigma_2, Q_1\times Q_2, \delta)$ where $\delta$ consists of:
\begin{center}
$\begin{array}{rl}
\multicolumn{2}{l}{\{\XStransition{(q,p)}{gg'|ff'}{(q',p')} \mid \XStransition{q}{g|f}{q'}\in\delta_1 \wedge \XStransition{p}{g'|f'}{p'}\in \delta_2\}}\\
\cup & \{\XStransition{(q,p)}{gp^\sharp|f}{(q',p)} \mid
\XStransition{q}{g|f}{q'}\in\delta_1 \wedge  p\in Q_2
\}\\
\cup& \{\XStransition{(q,p)}{gq^\sharp|f}{(q,p')} \mid
\XStransition{p}{g|f}{p'}\in\delta_2 \wedge q\in Q_1
\}
\end{array}$
\end{center}
\end{definition}

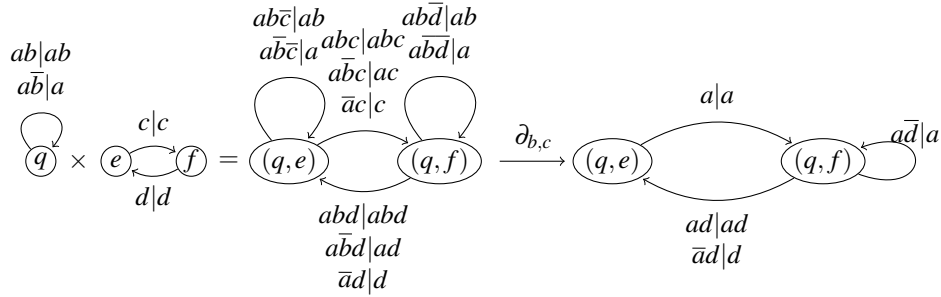
\begin{figure}[h]
\small
\begin{center}
\begin{tikzpicture}[shorten   >=1pt, auto]
         \tikzstyle{every state}=[inner sep=0pt,minimum size=6mm];
         \tikzstyle{initial}=[ellipse, inner sep=0pt,minimum size=6mm];
              \node[state] (q)     {$q$};
              \node at (.5,0)(t)     {$\times$ };
              \node[state] at (1,0) (e)           {$e$ };
              \node[state] (f) [right of=e]    {$f$ };
              \node at (2.5,0) (i)     {$=$ };
              \node[state, initial] at (3.3,0) (qe)           {$(q,e)$ };
              \node[state,initial] at (5.3,0) (qf)     {$(q,f)$ };
                   \path[->]
                   (q) edge [out=130,in=50,loop] node
{$\begin{array}{c}ab|ab\\a\overline
b|a\end{array}$} ()
                   (e) edge[bend left] node[above,sloped]  {$c|c$} (f)
                   (f) edge[bend left] node[below,sloped]  {$d|d$} (e)
                   (qe) edge[bend left] node[above,sloped]
                      {$\begin{array}{c}abc|abc\\a\overline b
c|ac\\ \overline a c|c \end{array}$} (qf)
                   (qf) edge[bend left] node[below,sloped]
{$\begin{array}{c}abd|abd\\a\overline b d|ad\\ \overline a d |d\end{array}$} (qe)
                   (qe) edge [out=130,in=50,loop] node
{$\begin{array}{c}ab\overline c|ab\\a\overline
b\overline c|a\end{array}$} ()
                   (qf) edge [out=130,in=50,loop] node
{$\begin{array}{c}ab\overline d|ab\\a\overline
b\overline d|a\end{array}$} ()
                    ;
		\draw[->] (6.1,0) to node[above] {$\partial_{b,c}$} (7,0);
		\node[state, initial] (qe2) [right of=qf,xshift=1.3cm]           {$(q,e)$ };
                \node[state,initial] (qf2) [right of=qe2,xshift=1.8cm]     {$(q,f)$ };
                   \path[->]
                   (qe2) edge[bend left] node[above,sloped]
                      {$\begin{array}{c}a|a%\\a\overline b c|ac
\end{array}$} (qf2)
                   (qf2) edge[bend left] node[below,sloped]
{$\begin{array}{c}ad|ad\\ \overline a d|d\end{array}$} (qe2)
                   (qf2) edge [out=-20,in=20,loop, above] node
{$\begin{array}{c}a\overline d|a\end{array}$} ()
                    ;
        \end{tikzpicture}
\end{center}
% \begin{comment}
% \begin{tikzpicture}[shorten   >=1pt, auto]
%          \tikzstyle{every state}=[inner sep=0pt,minimum size=6mm];
%                   \tikzstyle{initial}=[ellipse, inner sep=0pt,minimum
% size=6mm];
%           \node[state] (q1)     {$q_1$};
%           \node at (.5,0)(t)     {$\times$ };
%           \node[state] at (1,0) (q2)    {$q_2$ };
%            \node at (3.2,0) (i)     {$=$ };
%           \node[state,initial]  at (4,0) (q12)     {$(q_1,q_2)$ };
%
%
%                    \path[->]
%                    (q) edge [in=150,out=210,loop] node {$ab|ab$} ()
%                    (q2) edge [out=330,in=30,loop]    node[right]
% {$cd|cd, c\overline d |c$} ()
%                                       (q12) edge [out=345,in=15,loop]
% node[right] {$\begin{array}{c}abcd|abcd\\ abc\overline d|abc\\
% ab\overline c|ab\\ cd(\overline a\vee\overline b)|cd\\c\overline
% d(\overline a\vee\overline b)|c\end{array}$} ()
%
%                      ;
%         \end{tikzpicture}
% \end{comment}
\vspace{-4mm}
\caption{Product of {\sf LossySync} and {\sf FIFO1} and its synchronization of nodes $b$ and $c$}
\label{fig:product}
\end{figure}

\noindent
Here and throughout, we use $f\!f'$ as a shorthand for $f\cup f'$.
The first term in the union, above, applies when both automata fire in parallel.
The other terms apply when one automaton fires and the other is unable to (given by $p^\sharp$ and $q^\sharp$, respectively). Note that the product operation is closed for Reo automata, since it preserves
reactivity and uniformity \cite{BCS09}. Figure~\ref{fig:product} shows an example of the product of two automata.

We now define a synchronization operation that corresponds to joining two nodes
in a Reo connector. In order for this operation
to be well-defined we need that every guard in a transition label in
the automata is a conjunction of literals.
Note that in the automata presented in
Figure~\ref{basic_channels} for
basic Reo channels this is already the case, and moreover, it is always
possible to transform any guard $g$ into this form, by taking its
disjunctive normal form (DNF) $g_1\vee \ldots\vee g_k$ and splitting
the transition $g|f$ into the several $g_i|f$, for $i=1,\ldots,k$.
Given a transition relation $\delta$ we call $norm(\delta)$ the normalized
transition relation obtained from $\delta$ by putting all its guards in
DNF and splitting the transitions as explained above.

When
synchronizing two nodes $a$ and $b$ (which are then made internal), in
the resulting automaton, only the transitions where either both $a$ and $b$ or
neither $a$ nor $b$ fire are kept --- that is, $a$ and $b$ synchronize.
In order to propagate context information (requests), we require
that every guard contains either $a$ or $b$, expressed by
the condition $g\not\le\overline{a}\overline{b}$ below, which
more or less corresponds to an internal node acting like
a \emph{self-contained pumping station}~\cite{Arbab04}, meaning
that an internal node cannot store data or actively block behavior.

\begin{definition}\cite{BCS09} %[Synchronization]
Given a Reo automaton
$\mathcal{A}=(\Sigma, Q, \delta)$, we define the {\em synchronization} for $a,b\in\Sigma$ as
$\partial_{a,b}\mathcal{A}= (\Sigma, Q, \delta')$  where
\begin{center}
$\delta' =
\{\XStransition{q}{g\setminus_{ab}|f\setminus\{a,b\}}{q'} \mid
\XStransition{q}{g|f}{q'}\in norm(\delta) ~s.t.\ \ g\not\leq \overline a\overline b\ and\ a\in
f\Leftrightarrow b\in f\}$
\end{center}
\end{definition}

\noindent
Here and throughout, $g\setminus_{ab}$ is the guard obtained from $g$ by
deleting all
occurrences of $a$ and $b$. It is worth noting that synchronization
preserves reactivity and uniformity.

Figure \ref{fig:product} depicts the product of {\sf LossySync} and {\sf FIFO1}, together with the result of synchronizing nodes $b$ and $c$. This synchronized result provides the semantics for the {\sf LossyFIFO1} example in Figure~\ref{fig:lossyfifo}.

\subsubsection{Compositionality}
Given two Reo automata $\mathcal{A}_1$ and $\mathcal{A}_2$ over
the disjoint
alphabets $\Sigma_1$ and $\Sigma_2$, $\{a_1,\ldots, a_k\} \subseteq
\Sigma_1$ and $\{b_1,\ldots, b_k\}\subseteq\Sigma_2$ we construct
$
\partial_{a_1,b_1} \partial_{a_2,b_2} \cdots
\partial_{a_k,b_k}(\mathcal A_1\times\mathcal A_2)
$
as the automaton corresponding to a connector where node
$a_i$ of the first connector is connected to node $b_i$ of the
second connector,
for all $i\in\{1,\ldots,k\}$.
Note that the `plugging' order does not
matter because $\partial$ is commutative and it interacts well
with product.
These properties are captured in the following lemma.
\begin{lemma} \cite{BCS09} \label{prodandsync}
	For  the Reo automata
	$\mathcal{A}_1=(\Sigma_1, Q_1, \delta_1)$ and
	$\mathcal{A}_2=(\Sigma_2, Q_2, \delta_2)$:
	\begin{enumerate}
		\item  $\partial_{a,b}\partial_{c,d} \mathcal{A}_1=
			\partial_{c,d}\partial_{a,b} \mathcal{A}_1$, if $a,b,c,d\in\Sigma_1$.

		\item 	$\left(\partial_{a,b}\mathcal
			A_1\right)\times\mathcal A_2 \sim \partial_{a,b}(\mathcal A_1\times\mathcal A_2),$ if $a,b \notin \Sigma_2$
			%, if $a,b\in \Sigma_1$ and $\Sigma_1\cap\Sigma_2=\emptyset$.

%		\item $\partial_{a,c} (\mathcal {A}_1 \times Sync(a,b)) \sim \mathcal {A}_1 [b/c]$,
%		if $a,b\notin\Sigma_1$ and $c\in\Sigma_1$.
	\end{enumerate}
%	where $\mathcal{A} [b/c]$ is $\mathcal{A}$ with all occurrences of $c$
%	replaced by $b$.
\end{lemma}
\noindent
The notion of equivalence $\sim$ used above is bisimulation, defined
as follows.
\begin{definition}%[Bisimulation]
\cite{BCS09} Given the Reo automata ${\cal A}_1=(\Sigma, Q_1, \delta_1)$ and
${\cal
A}_2=(\Sigma, Q_2, \delta_2)$, we call $R\subseteq Q_1\times Q_2$ a
{\em bisimulation} iff for all $(q_1,q_2)\in R$:

If $\XStransition{q_1}{g|f}{q_1'}\in\delta_1$ and
$\alpha\in\at_\Sigma$, $\alpha \leq g$, then there exists a transition
$\XStransition{q_2}{g'|f}{q_2'}\in\delta_2$ such
that $\alpha\leq g'$ and $(q_1',q_2')\in R$ and vice-versa.
\end{definition}

\noindent
We say that two states $q_1\in Q_1$ and $q_2\in Q_2$ are bisimilar if
there exists a bisimulation relation containing the pair $(q_1,q_2)$ and we write
$q_1\sim q_2$. Two automata $\mathcal A_1$ and $\mathcal A_2$ are
bisimilar, written $\mathcal A_1\sim\mathcal A_2$, if there exists a bisimulation
relation such that every state of one automaton is related to some
state of the other automaton.

\section{Stochastic Reo}
\begin{figure}[h!]
\begin{center}
\begin{tikzpicture}[font=\tiny]
\node[point] (C) [label=below:$\gamma a$] {};
\node[point] (D) [right of=C,xshift=1cm,label=below:$\gamma b$] {};
\draw[sync] (C) to node[above] {$\gamma ab$} (D);

\node[point] (E) [right of =D, xshift=-0.8cm, label=below:$\gamma a$] {};
\node[point] (F) [right of=E,xshift=1cm,label=below:$\gamma b$] {};
\draw[lossysync] (E) to node[above] {$\gamma ab$} node[below,swap]
{$\gamma aL$} (F);

\node[point] (G) [right of =F, xshift=-0.8cm,label=below:$\gamma a$] {};
\node[point] (H) [right of=G,xshift=1cm,label=below:$\gamma b$] {};
\draw[syncdrain] (G) to node[above] {$\gamma ab$} (H);

\node[point] (A) [right of =H, xshift=-0.8cm, label=below:$\gamma a$,label=above right:$\gamma aF$] {};
\node[point] (B) [right of=A,xshift=1cm,label=below:$\gamma b$,label=above left:$\gamma Fb$] {};
\draw[fifo] (A) to (B);
\end{tikzpicture}
\end{center}
\vspace{-3mm}
\caption{Basic Stochastic Reo channels}
\label{fig:basicQchannel}
\end{figure}
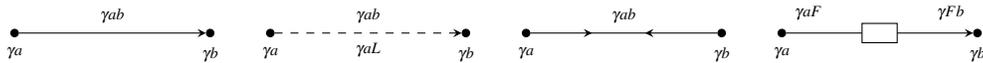

Stochastic Reo is an extension of Reo where channel ends and channels are annotated with stochastic values for \emph{data arrival rates} at channel ends and \emph{processing delay rates} at channels. Such rates are non-negative real values and describe how the probability that an event occurs varies with time. Figure~\ref{fig:basicQchannel}
shows the stochastic versions of the primitive Reo channels in Figure~\ref{fig:basic_channels}.
Here and throughout, for simplicity, we omit the node names, since they
can be inferred from the names of their respective arrival rates: for
instance, $\gamma a$ is the arrival rate of node $a$.

A processing delay rate represents how long it takes for a channel to perform
a certain activity, such as data-flow. For
instance, a {\sf LossySync} has two associated rates $\gamma ab$ and
$\gamma aL$ for, respectively, successful data-flow from node $a$ to node $b$, and losing the data item from node $a$.
In a {\sf FIFO1} $\gamma aF$ represents the delay for
data-flow from its source $a$ into the buffer, and $\gamma Fb$ for sending
the data from the buffer to the sink $b$.

Arrival rates describe the time between consecutive arrivals of I/O requests at
the source and sink nodes of Reo connectors. For instance, $\gamma a$ and
$\gamma b$ in Figure \ref{fig:basicQchannel} are the associated arrival rates
of write/take requests at the nodes $a$ and $b$.

Since arrival rates on nodes model their interaction with the environment only,
mixed nodes have no associated arrival rates. This is
justified by the fact that a mixed node delivers data items instantaneously to
the source end(s) of its connected channel(s). Hence, when joining a source with
a sink node into a mixed node, their arrival rates are discarded\footnote{For
simplicity, we assume ideal nodes whose activity incurs no delay. Any real
implementation of a node, of course, induces some processing delay rate. A real
node can be modeled as a composition of an ideal node with a {\sf Sync}
channel that manifests the processing delay rate. Thus, we can associate
delay distributions with Stochastic Reo nodes and automatically translate
them into such ``{\sf Sync} plus ideal node'' constructs .}.

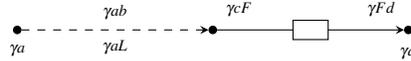
\begin{figure}[h]
\begin{center}
\begin{tikzpicture}[font=\tiny]
\node[point] (A) [label=below:$\gamma a$] {};
\node[point] (B) [right of=A,xshift=1cm, label=above right:$\gamma cF$] {};
% \node[point] (C) [right of=B,xshift=1cm, label=above right:$\gamma eF$, label=above left:$\gamma Fd$] {};
\node[point] (C) [right of=B,xshift=1cm, label=below:$\gamma d$, label=above left:$\gamma Fd$] {};
\draw[lossysync] (A) to node[above] {$\gamma ab$} node[below,swap] {$\gamma aL$} (B);
\draw[fifo]      (B) to (C);
% \node[point] (D) [right of=C,xshift=1cm,label=below:$\gamma g$, label=above left:$\gamma Fg$] {};
% \draw[fifo] (C) to (D);
\end{tikzpicture}
\end{center}
\vspace{-3mm}
\caption{Stochastic \sf{LossyFIFO1}}
\label{fig:stochasticlossyfifo}
\end{figure}

A stochastic version of the {\sf LossyFIFO1} is
depicted in Figure \ref{fig:stochasticlossyfifo}, including its
arrival and processing delay rates.

\subsection{Semantics: Stochastic Reo automata}

In this section, we provide a compositional semantics for
Stochastic Reo connectors, as an extension of Reo automata with functions that
assign stochastic values for data-flows and  I/O request arrivals.

\begin{definition}
A Stochastic Reo automaton is a triple $(\mathcal A, \mathbf r, \mathbf t)$ where $\mathcal A = (\Sigma, Q, \delta_{\mathcal A})$ is a Reo automaton and
\begin{itemize}
   \item $\mathbf r : \Sigma\to \mathbb R^{+}$ is a function that associates with each
   node its arrival rate.
   \item $\mathbf t: \delta_{\mathcal A} \to 2^\Theta$ is a function that associates
   with a transition a subset of $\Theta \subseteq 2^{\Sigma} \times 2^{\Sigma} \times
     \mathbb{R}^{+}$ such that each $(I,O,r)\in \Theta$ corresponds to a data-flow where $I$ is a set of input and/or mixed nodes; $O$ is a set of output and/or mixed nodes and $r$ is a processing delay rate for the data-flow.
\end{itemize}
\label{def:stochasticReoautomaton}
\end{definition}

\noindent
The Stochastic Reo automata corresponding to the {\sf LossySync} and {\sf FIFO1} in Figure \ref{fig:stochasticlossyfifo} are defined by the functions $\mathbf r$ and $\mathbf t$ shown in Table~\ref{tab:SreoAutomataExample}. Note that the function $\mathbf t$ is depicted in the transition, and function $\mathbf r$ is shown by a table.''. \vspace{-2mm}
\begin{table}[!h]
\begin{center}
\begin{small}
\begin{tabular}{|>{\centering}m{.25\linewidth} l | m{.25\linewidth}<{\centering} l|}
\hline
\begin{tikzpicture}[shorten   >=1pt, auto]
         \tikzstyle{every state}=[inner sep=0pt,minimum size=6mm];
         \tikzstyle{initial}=[ellipse, inner sep=0pt,minimum size=6mm];
              \node[state] (q) [yshift=-2cm]    {$q$};
	                	\path[->]
                   (q) edge [out=130,in=50,loop] node
{$\begin{array}{c}ab|ab,~\{(\{a\},\{b\},\gamma ab)\} \\
a\overline b|a,~\{(\{a\},\emptyset,\gamma aL)\} \end{array}$} ();
\end{tikzpicture}
&
\begin{tabular}{|c|c|}
\hline
& $r$ \\
\hline
$a$ & $\gamma a$\\
$b$ & $\gamma b$\\
\hline
\end{tabular}
&
\begin{tikzpicture}[shorten   >=1pt, auto]
         \tikzstyle{every state}=[inner sep=0pt,minimum size=6mm];
         \tikzstyle{initial}=[ellipse, inner sep=0pt,minimum size=6mm];
  	      \node[state] at (1,0) (e)           {$e$ };
              \node[state] (f) [right of=e,xshift=1cm]    {$f$ };
                   \path[->]
                   (e) edge[bend left] node[above,sloped]  {$c|c,~\{(\{c\},\emptyset,\gamma cF)\}$} (f)
                   (f) edge[bend left] node[below,sloped]  {$d|d,~\{(\emptyset,\{d\},\gamma Fd)\}$} (e);
\end{tikzpicture}
&
\begin{tabular}{|c|c|}
\hline
& $r$ \\
\hline
$c$ & $\gamma c$\\
$d$ & $\gamma d$\\
\hline
\end{tabular}\\

\hline
\end{tabular}
\end{small}
\end{center} \vspace{-1mm}
\caption{Stochastic Reo automaton for {\sf LossySync} and {\sf FIFO1}}
\label{tab:SreoAutomataExample}
\end{table}

An element of $\theta \in \Theta$ is
accessed by projection functions $i:\Theta \to 2^{\Sigma}$, $o:\Theta \to 2^{\Sigma}$ and $v:\Theta \to
\mathbb{R}^{+}$; $i(\theta)$ and $o(\theta)$ return, respectively, relevant input and output
nodes of a data-flow, and $v(\theta)$ returns the delay rate of a data-flow through nodes in $i(\theta)$ and $o(\theta)$.
%that characterizes the delay rate of a data-flow.

\begin{definition}
Given two Stochastic Reo automata $(\mathcal A_1,
\mathbf r_1, \mathbf t_1)$ and $(\mathcal A_2, \mathbf r_2, \mathbf t_2)$, their product is defined as $(\mathcal A_1, \mathbf r_1, \mathbf t_1)\times (\mathcal A_2, \mathbf r_2, \mathbf t_2)=(\mathcal A_1
\times \mathcal A_2, \mathbf r_1 \cup \mathbf r_2, \mathbf t)$ where
\begin{center}
$\begin{array}{@{}l}
\mathbf t(\XStransition{(q,p)}{gg'|ff'}{(q',p')}) =\mathbf t_1(\XStransition{q}{g|f}{q'}) \cup \mathbf t_2(\XStransition{p}{g'|f'}{p'})\\
\mathbf t(\XStransition{(q,p)}{g|f}{(q',p)}) =
\mathbf t_1(\XStransition{q}{g|f}{q'})\\
\mathbf t(\XStransition{(q,p)}{g'|f'}{(q,p')}) =
\mathbf t_2(\XStransition{p}{g'|f'}{p'})\\
\end{array}$
\end{center}
\end{definition}
\noindent
Note that we use $\times$ to denote both the product of Reo automata and the product of Stochastic Reo automata.

The set of 3-tuples that $\mathbf t$ associates with a transition $m$ represents the composition of the delay rates involved in all data-flows synchronized by the transition $m$. In order to keep Stochastic Reo automata generally useful and compositional, and their product commutative, we avoid fixing the precise formal meaning of distribution rates of synchronized transitions composed in a product; instead, we present the ``delay rate'' of their composite transition in the product automaton as the union of the delay rates of the synchronizing transitions of the two automata. How exactly these rates combine to yield the composite rate of the transition depends on different properties of the distributions and their time ranges. For example, in the continuous-time case, no two events can occur at the same time; whereas the exponential distributions are not closed under taking maximum. In Section~\ref{sec:translation} we show  how to translate a Stochastic Reo automaton to a CTMC by the union of rates of the exponential distribution in the continuous-time case.

\begin{definition}
For a Stochastic Reo automaton $(\mathcal A,\mathbf r, \mathbf t)$, the synchronization operation on nodes $a$ and $b$ is defined as  $\partial_{a,b}(\mathcal A,\mathbf r,\mathbf t)=(\partial_{a,b}\mathcal{A}, \mathbf r', \mathbf t')$ where
\begin{itemize}
 \item $\mathbf r'$ is $\mathbf r$ restricted to the domain $\Sigma\setminus\{a,b\}$.
 \item $\mathbf t'$ is defined as:
\begin{center}
$\begin{array}{ll}
\multicolumn{2}{l}{\mathbf t'(\XStransition{q}{g\setminus_{ab}|f\setminus\{a,b\}}{q'})= \{(A',B',r) \mid (A,B,r) \in \mathbf t(\XStransition{q}{g|f}{q'}),~~~~~~~~~~~~~~~~} \\ \multicolumn{2}{r}{A'=sync(A,\{a,b\}) \wedge~ B'=sync(B,\{a,b\})}\}
\end{array}$
\end{center}
 \item $sync: 2^{\Sigma} \times 2^{\Sigma} \to 2^{\Sigma}$ gathers nodes  joined by synchronization, and is defined as:
\begin{center}
$sync(A,B)
= \left \{ \begin{array}{ll}
             A \cup B & \qquad \textrm{if } A \cap B \neq \emptyset
             \\ A & \qquad \textrm{otherwise}\\
          \end{array}\right.$
\end{center}
\end{itemize}
\end{definition}

\noindent
Note that we use the symbol $\partial_{a,b}$ to denote both the synchronization of Reo automata and the synchronization of Stochastic Reo automata.

We now revisit the {\sf LossyFIFO1} example. Its semantics is given by the
triple $(\mathcal A_{LossyFIFO1}, \mathbf r, \mathbf t)$, where $\mathcal A_{LossyFIFO1}$ is the automaton depicted in
Figure~\ref{fig:product} and $\mathbf r$ is defined as $\mathbf r = \{a
\mapsto \gamma a , d \mapsto \gamma d  \}$. For $\mathbf t$, we first compute
$\mathbf t_{LossySync\times FIFO1}$:\\
\begin{small}
\begin{tabular}{>{\centering}m{.5\linewidth} m{.3\linewidth}<{\centering}}
 \begin{tikzpicture}[shorten   >=1pt, auto]
         \tikzstyle{every state}=[inner sep=0pt,minimum size=6mm];
         \tikzstyle{initial}=[ellipse, inner sep=0pt,minimum size=6mm];
              \node[state, initial] at (1.3,0) (qe)           {$(q,e)$ };
              \node[state,initial] at (6.3,0) (qf)     {$(q,f)$ };
                   \path[->]
                   (qe) edge[bend left] node[above,sloped]
                      {$\begin{array}{c}{\bf abc|abc,}~\boldsymbol{\Theta_3}\\
a\overline bc|ac,~\Theta_4\\
c\overline a |c,~\Theta_5
\end{array}$} (qf)
                   (qf) edge[bend left] node[below,sloped]
{$\begin{array}{c}
abd|abd,~\Theta_6\\
{\bf a\overline b d|ad,}~\boldsymbol{\Theta_7}\\
{\bf d\overline a |d,}~\boldsymbol{\Theta_8}
\end{array}$} (qe)
                   (qe) edge [out=130,in=50,loop] node[yshift=4]
{$\begin{array}{c}
ab\overline c|ab,~\Theta_1\\
a\overline b\overline c|a,~\Theta_2
\end{array}$} ()
                   (qf) edge [out=130,in=50,loop] node
{$\begin{array}{c}
ab\overline d|ab,~\Theta_1\\
{\bf a\overline b\overline d|a,}~\boldsymbol{\Theta_2}
\end{array}$} ()
                    ;
        \end{tikzpicture}
&
\begin{tabular}{|ll|}
\hline
$\Theta_1:$ & $\{(\{a\},\{b\},\gamma ab)\}$\\
$\Theta_2:$ & $\{(\{a\},\emptyset,\gamma aL)\}$\\
$\Theta_3:$ & $\{(\{a\},\{b\},\gamma ab),(\{c\},\emptyset,\gamma cF)\}$\\
$\Theta_4:$ & $\{(\{a\},\emptyset,\gamma aL),(\{c\},\emptyset,\gamma cF)\}$\\
$\Theta_5:$ & $\{(\{c\},\emptyset,\gamma cF)\}$\\
$\Theta_6:$ & $\{(\{a\},\{b\},\gamma ab),(\emptyset,\{d\},\gamma Fd)\}$\\
$\Theta_7:$ & $\{(\{a\},\emptyset,\gamma aL),(\emptyset ,\{d\},\gamma Fd)\}$\\
$\Theta_8:$ & $\{(\emptyset,\{d\},\gamma Fd)\}$\\
\hline
\end{tabular}
\end{tabular}
\end{small}

\noindent
Above, the labels that correspond to the
transitions that will be kept after synchronization appear in {\bf bold}.
Thus, the result of joining  nodes by synchronization, is shown in Figure~\ref{fig:SRAlossyfifo} as:
\vspace{-3mm}
\begin{small}
\begin{figure}[!h]
\begin{center}
 \begin{tikzpicture}[shorten   >=1pt, auto]
         \tikzstyle{every state}=[inner sep=0pt,minimum size=6mm];
         \tikzstyle{initial}=[ellipse, inner sep=0pt,minimum size=6mm];
		\node[state, initial] at (1.3,0) (qe2) {$(q,e)$};
                \node[state,initial] at (5.3,0) (qf2) {$(q,f)$};
                   \path[->]
                   (qe2) edge[bend left] node[above,sloped]
                      {$\begin{array}{c}a|a\\
\{(\{a\},\{{\bf b},{\bf c}\},\gamma ab),(\{{\bf b},{\bf c}\},\emptyset,\gamma cF)\}%\\a\overline b c|ac
\end{array}$} (qf2)
                   (qf2) edge[bend left] node[below,sloped]
{$\begin{array}{ll}ad|ad, & \{(\{a\},\emptyset,\gamma aL),(\emptyset ,\{d\},\gamma Fd)\}\\
d\overline a|d, & \{(\emptyset,\{d\},\gamma Fd)\}
\end{array}$} (qe2)
                   (qf2) edge [out=-20,in=20,loop, above] node [right]
{$\begin{array}{c}a\overline d|a,~\{(\{a\},\emptyset,\gamma aL\end{array})\}$} ()
                    ;
\end{tikzpicture}
\end{center}
\vspace{-4mm}
\caption{Stochastic Reo automaton for {\sf LossyFIFO1}}
\label{fig:SRAlossyfifo}
\end{figure}
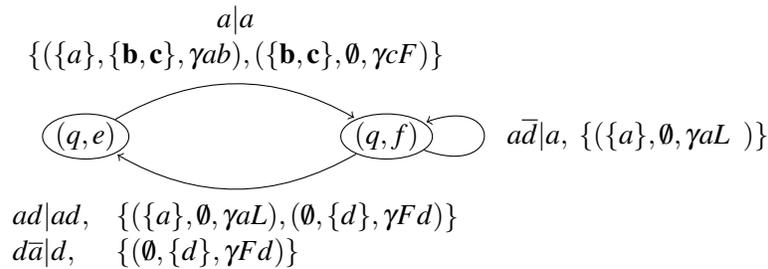
\end{small}

\noindent
Note that the port names that appear in {\bf bold} represent the synchronization of nodes $b$ and $c$.

In this way, we can carry in the semantic model of Reo circuits, given as Reo
automata, stochastic information, i.e., arrival rates and processing
delay rates that pertain to its QoS.

Definition \ref{def:stochasticReoautomaton} shows that our extension of Reo automata deals with such stochastic information separately, apart from the underlying Reo automaton. Thus, our extended model retains the properties of Reo automata, i.e., the compositionality result presented in Section 2.1.2 can be extended to Stochastic Reo automata:
\begin{lemma}
For two disjoint Stochastic Reo automata $(\mathcal A_1, \mathbf r_1, \mathbf t_1)$ and $(\mathcal A_2, \mathbf r_2, \mathbf t_2)$ with ${\cal A}_1=(\Sigma, Q_1, \delta_1)$ and ${\cal A}_2=(\Sigma, Q_2, \delta_2)$,
\begin{enumerate}
\item $\partial_{a,b} \partial_{c,d} (\mathcal A_1, \mathbf r_1, \mathbf t_1) = \partial_{c,d} \partial_{a,b} (\mathcal A_1, \mathbf r_1, \mathbf t_1)$,  if $a,b,c,d\in\Sigma_1$
\item $(\partial_{a,b} (\mathcal A_1, \mathbf r_1, \mathbf t_1) )\times (\mathcal A_2, \mathbf r_2, \mathbf t_2) \sim \partial_{a,b} ((\mathcal A_1, \mathbf r_1, \mathbf t_1) \times (\mathcal A_2, \mathbf r_2, \mathbf t_2))$, if $a,b \notin \Sigma_2$
\end{enumerate}
\end{lemma}
\noindent
Here $(\mathcal A_1, \mathbf r_1, \mathbf t_1) \sim (\mathcal A_2, \mathbf r_2, \mathbf t_2)$  if and only if $\mathcal A_1\sim \mathcal A_2$, $\mathbf r_1=\mathbf r_2$ and $\mathbf t_1=\mathbf t_2$. Because of space limitation, we leave the proof of this lemma for an extended version of this paper.
% Note that for the reasons of space, the proof for this lemma will be shown in an extended version of this paper.

\section{Translation to CTMC} \label{sec:translation}
In this section, we show how to translate a Stochastic Reo automaton into a homogeneous CTMC model. A homogeneous CTMC is a stochastic process with 1) homogeneity, 2) memoryless/Markov property, and 3) discrete state space in the continuous-time domain~\cite{IMC02}. These properties yield efficient methodologies for numerical analysis.

In the continuous-time domain, the exponential distribution is the only one that satisfies the memoryless  property. Therefore, for the translation, we assume that the rates of data-arrivals and data-flows are exponentially distributed.

A CTMC model derived from a Stochastic Reo automaton
$(\mathcal A,\mathbf r,\mathbf t)$ with $\mathcal A = (\Sigma, Q, \delta_{\mathcal A})$ is a
pair $(S,\delta)$ where $S=S_A \cup S_M$ is the set of states. $S_A$ represents the configurations of the
system derived from its Reo automaton and the pending status of I/O requests; $S_M$ is the set of states that result from the micro-step division of synchronous actions~(see below). $\delta = \delta_{Arr} \cup \delta_{Proc}
\subseteq S \times \mathbb{R}^{+} \times S$, explained below, is the set of transitions, each
labeled with a stochastic value specifying the arrival or the processing delay rate of the transition. $\delta_{Arr}$ and $\delta_{Proc}$ are defined in Section 4.3.

A state in $S$ models a configuration of the connector,
including the presence of the I/O requests pending on its boundary nodes, if
any. Data-arrivals change system configuration only by changing the pending
status of their respective boundary nodes. Data-flows corresponding to a
transition of a Reo automaton change the system configuration, and release the
pending I/O requests on its involved boundary nodes.

In a CTMC model, the probability that two events (e.g., the arrival
of an I/O request, the transfer of a data item, a processing step, etc.) happen at the same time is \emph{zero}: only a single event occurs at a time. In compliance
with this requirement, for a Stochastic Reo automaton
$(\mathcal A,\mathbf r,\mathbf t)$ with $\mathcal A=(\Sigma,Q,\delta_{\mathcal A})$ and a set
of boundary nodes $\Sigma' \subseteq \Sigma$, the set $S_A$ and the preliminary set of data-arrival transitions of the CTMC derived for $(\mathcal A,\mathbf r,\mathbf t)$  are defined as:

\begin{center}
$\begin{array}{lll}
S_A & = & \{(q,R)
\mid
q \in Q,~R \subseteq \Sigma' \} \\
\delta'_{Arr} & = & \{\XStransition{(q,R)}{\mathbf r (c)}{(q,R \cup \{c\})}
 \mid
(q,R),~(q,R \cup \{c\}) \in S_A,~c \notin R\}
\end{array}$
\end{center}
\noindent
The set $\delta'_{Arr}$ is used in Section~\ref{sec:deriving} to define $\delta_{Arr}$.

\subsection{Micro-step transitions}
The CTMC transitions associated with data-flows are more complicated
since groups of synchronized data-flows are modeled as a single transition in a Reo automaton. Therefore, we need to divide such synchronized data-flows into so-called micro-step transitions, respecting the connection information, i.e., the
topology of a Reo connector. %Note that this division delineates synchronized data-flows, and does not divide each data-flow itself.

The connection information can be recovered from the
3-tuples associated with each transition in a Reo automaton since the first and the second elements of a 3-tuple describe the input and the output nodes, respectively, involved in the data-flow of its transition,
and the data-flow in the transition occurs from its input to its output nodes.

For example, the transition $\XStransition{(q,e)}{a|a}{(q,f)}$ in the Reo automaton of the {\sf LossyFIFO1} example in Figure~\ref{fig:SRAlossyfifo} has a set of the 3-tuples $\{(\{a\},\{b,c\},\gamma ab),(\{b,c\},\emptyset,\gamma cF)\}$. The connection information inferred from this set states that data-flow occurs from $a$ to the buffer
through $b$ and $c$. The transition is thus divided into two consecutive micro-step
transitions $(\{a\},\{b,c\},\gamma ab)$ and
$(\{b,c\},\emptyset,\gamma cF)$.

Such data-flow information of each transition in a Reo automaton is formalized by a \emph{delay-sequence} defined by the following grammar:
\begin{center}
$\Lambda \ni \lambda ::= \epsilon \mid \theta \mid \lambda | \lambda \mid \lambda
;\lambda$
\end{center}
where $\epsilon$ is the empty sequence and $\theta$ is a 3-tuple $(I,O,r)$ for a
primitive Reo channel. $\lambda | \lambda$ denotes
parallel composition, and $\lambda ; \lambda$ denotes sequential
composition. The empty sequence $\epsilon$ is an identity element for $;$ and $|$, $|$ is commutative, associative, and
idempotent, $;$ is associative and distributes over $|$.

\subsection{Extracting delay-sequences}
The delay-sequence corresponding to a set of 3-tuples associated with a transition in a Stochastic Reo automaton is obtained by Algorithm \ref{alg:extract}.1. Note that if the parameter of the function \textbf{Ext} is a singleton, then $\textbf{Ext}(\{\theta\})=\theta$ since $i(\theta) \cap o(\theta)=\emptyset$.

\begin{algorithm} \textbf{Ext($\Theta$)} where
$\Theta=\mathbf t(\XStransition{p}{g|f}{q})$
\label{alg:gettingDS}
\begin{algorithmic}
\STATE $S=\epsilon,~\mathit{toGo}=\Theta,$ $Init:=\{\theta \in \Theta
\mid
i(\theta) \cap o(\theta')=\emptyset ~\textrm{for all } \theta' \in \Theta\}$
\FOR{$\theta \in Init$}
 \STATE $\lambda_{\theta}:= \theta,~Pre :=  \{\theta\}, ~\mathit{toGo} := \mathit{toGo}
 \setminus Pre$
 \STATE $Post = \{\theta \in \mathit{toGo}
 \mid
 \exists \theta' \in Pre ~s.t.~o(\theta') \cap i(\theta) \neq \emptyset\}$
 \WHILE{$Post \neq \emptyset$}
  \STATE $\lambda' := (\theta_1| \cdots | \theta_k)$ where $
  Post=\{\theta_1,\cdots,\theta_k\}$
  \STATE $\lambda_{\theta} :=\lambda_{\theta} ; \lambda',~Pre := Post,~ \mathit{toGo}
  :=\mathit{toGo} \setminus Pre$
  \STATE $Post := \{\theta \in \mathit{toGo}
 \mid
 \exists \theta' \in Pre ~s.t. ~o(\theta') \cap i(\theta) \neq \emptyset \}$
 \ENDWHILE
 \STATE $S:= S | \lambda_{\theta}$
\ENDFOR
\STATE \textbf{return} S
\end{algorithmic}
\label{alg:extract}
\caption{Extraction of a delay-sequence out of a set $\Theta$ of 3-tuples}
\end{algorithm}

Intuitively, the \textbf{Ext} function delineates the set of activities that -- at the level of a Stochastic Reo automaton -- must happen synchronously/atomically, into corresponding delay-sequences. If a certain data-flow associated with a 3-tuple $\theta_1$ explicitly
precedes another one $\theta_2$, then $\theta_1$ is sequenced before
$\theta_2$, i.e., encoded as $\theta_1;\theta_2$. Otherwise, they can occur in any order, encoded as $\theta_1 | \theta_2$.

Applying Algorithm \ref{alg:extract}.1 to the {\sf LossyFIFO1} example yields the following result:\\
\begin{small}
\begin{tabular}{>{\centering}m{.5\linewidth} m{.3\linewidth}<{\centering}}
 \begin{tikzpicture}[shorten   >=1pt, auto]
         \tikzstyle{every state}=[inner sep=0pt,minimum size=6mm];
         \tikzstyle{initial}=[ellipse, inner sep=0pt,minimum size=6mm];
          \node[state, initial] at (1.3,0) (qe2) {$(q,e)$};
                \node[state,initial] at (5.3,0) (qf2) {$(q,f)$};
                   \path[->]
                   (qe2) edge[bend left] node[above,sloped]
                      {$\begin{array}{c}a|a,~\lambda_1%\\a\overline b c|ac
\end{array}$} (qf2)
                   (qf2) edge[bend left] node[below,sloped]
{$\begin{array}{l}ad|ad,~\lambda_3\\
\overline a d|d, ~\lambda_4
\end{array}$} (qe2)
                   (qf2) edge [out=-20,in=20,loop, above] node
{$\begin{array}{c}~~~~~~a\overline d|a,~\lambda_2 \end{array}$} ()
                    ;
        \end{tikzpicture}
&
\begin{tabular}{|ll|}
\hline
$\lambda_1:$ & $(\{a\},\{b,c\},\gamma ab) ~{\bf ;}~ (\{b,c\},\emptyset,\gamma cF)$\\
$\lambda_2:$ & $(\{a\},\emptyset,\gamma aL)$\\
$\lambda_3:$ & $(\{a\},\emptyset ,\gamma aL) ~{\bf |}~ (\emptyset ,\{d\},\gamma
Fd)$\\
$\lambda_4:$ & $(\emptyset,\{d\},\gamma Fd)$\\
\hline
\end{tabular}
\end{tabular}
\end{small}

The parameter $\Theta$ of Algorithm \ref{alg:extract}.1 is a finite set
of 3-tuples, and $\mathit{Init}$, $\mathit{Post}$ and $\mathit{toGo}$, subsets of $\Theta$, are also finite. Moreover,
$\mathit{Post}$ becomes eventually $\emptyset$ since $\mathit{toGo}$ decreases during the procedure.
Thus, we can conclude that Algorithm \ref{alg:extract}.1 always
terminates.

\subsection{Deriving the CTMC} \label{sec:deriving}
We now show how to derive the transitions in the CTMC model from the transitions in a stochastic Reo automaton. We do this in two steps:
\begin{enumerate}
 \item For each transition $\XStransition{p}{g|f}{q}\in \delta_{\mathcal A}$, we derive transitions $\XStransition{(p,R)}{\lambda~~}{(q,R\setminus f)}$ for every set of pending requests $R$ that suffices to activate the guard $g$ ($\widehat{R}\leq g \setminus \widehat{\overline{\Sigma}}$), where $\lambda$ is the delay-sequence associated with the set of 3-tuples $\mathbf t(\XStransition{p}{g|f}{q})$. This set of derived transitions is defined below as $\delta_{Macro}$.
 \item We divide a transition in $\delta_{Macro}$ labeled by $\lambda$ into a combination of micro-step transitions, each of which corresponds to a single event.
\end{enumerate}
The following figure briefly illustrates the procedure mentioned above:
\begin{center}
\begin{tabular}{|>{\centering}m{.4\linewidth}|m{.4\linewidth}<{\centering}|}
\hline
 $\XStransition{p}{\lambda_1 ; \lambda_2}{q}$ & $\XStransition{p}{\lambda_1 | \lambda_2}{q}$\\
\hline
\begin{tikzpicture}
 \node[state] (A)  {$p$};
 \node[state] (B) [right of=A] {$s_1$};
 \node[state] (D) [right of=B,xshift=8mm] {$s_i$};
 \node[state] (E) [right of=D] {$s_k$};
 \node[state] (C) [right of=E,xshift=8mm] {$q$};
 \draw[transition] (A) to (B);
 \draw[->,>=stealth,dotted] (B) to (D);
 \draw[transition] (D) to (E);
 \draw[->,>=stealth,dotted] (E) to (C);
 \draw[decorate,decoration={brace,raise=7pt}] (A) to node[above,yshift=8pt] {$\lambda_1$} (D);
 \draw[decorate,decoration={brace,raise=7pt}] (D) to node[above,yshift=8pt] {$\lambda_2$} (C);
\end{tikzpicture}
&
\begin{tikzpicture}
 \node[state] (a) [yshift=-15mm]{$p$};
 \node[state] (b) [right of=a,yshift=3.5mm,xshift=-2mm] {$s_1$};
 \node[state] (c) [right of=a,yshift=-3.5mm,xshift=-2mm] {$s_2$};
 \node[state] (e) [right of=b,yshift=6.5mm,xshift=5mm] {$s_i$};
 \node[state] (f) [right of=c,yshift=-6.5mm,xshift=5mm] {$s_j$};
 \node[state] (g) [right of=e,yshift=-3.5mm,xshift=-2mm] {$s_k$};
 \node[state] (h) [right of=f,yshift=3.5mm,xshift=-2mm] {$s_l$};
 \node[state] (i) [right of=a,xshift=7mm] {$s_3$};
%  \node[ellipse,draw,inner sep=0pt,minimum size=4mm] (i) [right of=a,xshift=7mm] {$s_{1,2}$};
 \node[state] (d) [right of=a,xshift=3.5cm] {$q$};
 \draw[transition] (a) to (b);
 \draw[transition] (a) to (c);
 \draw[->,>=stealth,dotted] (b) to (e);
 \draw[->,>=stealth,dotted] (c) to (f);
 \draw[transition] (e) to (g);
 \draw[transition] (f) to (h);
 \draw[->,>=stealth,dotted] (g) to (d);
 \draw[->,>=stealth,dotted] (h) to (d);
 \draw[transition] (b) to (i);
 \draw[transition] (c) to (i);
 \draw[->,>=stealth,dotted] (i) to (g);
 \draw[->,>=stealth,dotted] (i) to (h);
 \draw[decorate,decoration={brace,raise=8pt}] (a) to node[above,yshift=8pt] {$\lambda_1$} (e);
 \draw[decorate,decoration={brace,raise=8pt}] (e) to node[above,yshift=8pt] {$\lambda_2$} (d);
%  \draw[decorate,decoration={brace,raise=-8pt}] (a) to node[below,yshift=-8pt] {$\lambda_1$} (f);
 \draw[decorate,decoration={brace,raise=8pt,mirror}] (a) to node[below,yshift=-8pt] {$\lambda_2$} (f);
 \draw[decorate,decoration={brace,raise=8pt,mirror}] (f) to node[below,yshift=-8pt] {$\lambda_1$} (d);
%  \draw[transition] (b) to node[above] {$\lambda_2$} (d);
%  \draw[transition] (c) to node[below] {$\lambda_1$} (d);
%  \draw[transition] ($(a)!.25!(c)$) to ($(b)!.25!(d)$);
%  \draw[transition] ($(a)!.25!(b)$) to ($(c)!.25!(d)$);
%  \draw[loosely dotted] ($(a)!.65!(c)$) to ($(b)!.65!(d)$);
%  \draw[loosely dotted] ($(a)!.65!(b)$) to ($(c)!.65!(d)$);
 \end{tikzpicture}
\\
\hline
\end{tabular}
\end{center}
\noindent
A sequential delay-sequence $\lambda_1 ; \lambda_2$ allows for the events corresponding to $\lambda_1$ to occur before the ones corresponding to $\lambda_2$. For a parallel delay-sequence $\lambda_1 | \lambda_2$, events corresponding to $\lambda_1$ and $\lambda_2$ occur in an interleaving way, while they preserve their respective order of occurrence in $\lambda_1$ and $\lambda_2$. All indexed states $s_n$ are included in $S_M$ which consists of the states derived from the division of synchronized data-flows into micro-step transitions.

Given a Stochastic Reo automaton $(\mathcal A,\mathbf r,\mathbf t)$ with
$\mathcal A=(\Sigma,Q,\delta_{\mathcal A})$ and a set of boundary nodes $\Sigma'$, a macro-step transition relation for the
synchronized data-flows is defined as:
% \begin{center}
% $\begin{array}{ll}
% \multicolumn{2}{l}{\delta_{Macro}=\{\XStransition{(p,R)}{\lambda~~~}{(q,R\setminus f)}
% \mid
% \XStransition{p}{g|f}{q} \in \delta_{\mathcal A}, R \subseteq \Sigma',~ \widehat{R}\leq g \setminus \widehat{\overline{\Sigma}},~~~~}\\
% \multicolumn{2}{r}{\lambda=\textbf{Ext}(\mathbf t(\XStransition{p}{g|f}{q}))\}}
% \end{array}$
% \end{center}
\begin{center}
$
 \delta_{Macro}=\{\XStransition{(p,R)}{\lambda}{(q,R\setminus f)}~
\mid~
\XStransition{p}{g|f}{q} \in \delta_{\mathcal A},~ R \subseteq \Sigma',~ \widehat{R}\leq g \setminus \widehat{\overline{\Sigma}},~\lambda=\textbf{Ext}(\mathbf t(\XStransition{p}{g|f}{q}))\}
$
\end{center}

We explicate a macro-step transition with a number of micro-step transitions, each of which
corresponds to a single data-flow. This refinement yields auxiliary states between the
source and the target states of the macro-step transition. Let $(p,R)$ be a source state for a data-flow
corresponding to a 3-tuple $\theta$. Then the generated auxiliary states are defined
as $(p_{\theta},R \setminus nodes(\theta))$ where $p_{\theta}$ is just a label denoting that
data-flows corresponding to $\theta$ have occurred, and the function $nodes: \Lambda \to 2^{\Sigma}$ is defined as:
\begin{center}
$\begin{array}{l}
nodes(\lambda) =\left \{ \begin{array}{ll} i(\theta) \cup o(\theta) & \qquad
\textrm{if } \lambda=\theta\\
nodes(\lambda_1) \cup nodes(\lambda_2) & \qquad \textrm{if }
\lambda=\lambda_1;\lambda_2~\vee \lambda=\lambda_1 | \lambda_2 \end{array}
\right.
\end{array}$
\end{center}
\noindent
The set of such auxiliary states is obtained as
$S_M = states(\XStransition{(p,R)}{\lambda}{(q,R')})$ where
% \begin{center}
% $\begin{array}{l}
% \mathit{states}(\XStransition{(p,R)}{\lambda}{(q,R')})=\\
% \left \{ \begin{array}{ll}
%         \{(p,R),(q,R')\} & \qquad \textrm{if } \lambda=\theta \\
%         \bigcup \mathit{states}(m) ~\forall m \in div(\XStransition{(p,R)}{\lambda}{(q,R')}) & \qquad \textrm{otherwise}
%          \end{array} \right.
% \end{array}$
% \end{center}
\[
\mathit{states}(\XStransition{(p,R)}{\lambda}{(q,R')})= \left \{ \begin{array}{ll}
        \{(p,R),(q,R')\} & \qquad \textrm{if } \lambda=\theta \\
        \bigcup \mathit{states}(m) ~\forall m \in div(\XStransition{(p,R)}{\lambda}{(q,R')}) & \qquad \textrm{otherwise}
         \end{array} \right.
\]
\noindent
The function $div: \delta_{Macro} \to 2^{\delta_{Macro}}$ is defined as:
% \begin{center}
% $\begin{array}{l}
% div(\XStransition{(p,R)}{\lambda}{(q,R')})= \\
%  \left \{ \begin{array} {lll}
%             \{\XStransition{(p,R)}{\theta}{(q,R')}\} &\multicolumn{2}{l}{\qquad \textrm{if } \lambda = \theta ~\wedge~ \nexists
%             \XStransition{(p,R)}{\theta}{(p',R')} \in \delta_{Macro}}\\
%             \multicolumn{2}{l}{div(\XStransition{(p,R)}{\lambda_1~~}{(p_{\lambda_1},R'')}) \cup div(\XStransition{(p_{\lambda_1},R'')}{~~\lambda_2}{(q,R')})} & \\
%             & \multicolumn{2}{l}{\qquad \textrm{if } \lambda=\lambda_1 ; \lambda_2~\textrm{where}~ R''=R \setminus nodes(\lambda_1)}\\
%             \multicolumn{2}{l}{\{m_1 \bowtie m_2 \mid m_i \in div(\XStransition{(p,R)}{\lambda_i~~}{(p_{\lambda_i},R'')}), ~i\in \{1,2\}\}} & \\
%             & \multicolumn{2}{l}{\qquad \textrm{if } \lambda = \lambda_1 | \lambda_2 ~\textrm{where } R''=R\setminus nodes(\lambda_i)} \\
%             \emptyset~~~~~~~~~~~~~~~~~~~~~~~~~~~~~~~~~~ & \multicolumn{2}{l}{\qquad \textrm{otherwise}} \end{array}
%             \right.
% \end{array}$
% \end{center}
\[
 div(\XStransition{(p,R)}{\lambda}{(q,R')})=  \left \{ \begin{array} {lll}
            \{\XStransition{(p,R)}{\theta}{(q,R')}\} &\multicolumn{2}{l}{\qquad \textrm{if } \lambda = \theta ~\wedge~ \nexists
            \XStransition{(p,R)}{\theta}{(p',R')} \in \delta_{Macro}}\\
            \multicolumn{2}{l}{div(\XStransition{(p,R)}{\lambda_1~~}{(p_{\lambda_1},R'')}) \cup div(\XStransition{(p_{\lambda_1},R'')}{~~\lambda_2}{(q,R')})} & \\
            & \multicolumn{2}{l}{\qquad \textrm{if } \lambda=\lambda_1 ; \lambda_2~\textrm{where}~ R''=R \setminus nodes(\lambda_1)}\\
            \multicolumn{2}{l}{\{m_1 \bowtie m_2 \mid m_i \in div(\XStransition{(p,R)}{\lambda_i~~}{(p_{\lambda_i},R'')}), ~i\in \{1,2\}\}} & \\
            & \multicolumn{2}{l}{\qquad \textrm{if } \lambda = \lambda_1 | \lambda_2 ~\textrm{where } R''=R\setminus nodes(\lambda_i)} \\
            \emptyset~~~~~~~~~~~~~~~~~~~~~~~~~~~~~~~~~~ & \multicolumn{2}{l}{\qquad \textrm{otherwise}} \end{array}
            \right.
\]
\noindent
where the function $\bowtie$ computes all interleaving compositions of the two
 transitions as: for every
$(p,R_1) \in states(\XStransition{s_2}{\theta_2}{s'_2})$ and for every $(p,R_2) \in states(\XStransition{s_1}{\theta_1}{s'_1})$
\begin{center}
$\begin{array}{lll}
\begin{array}{l c l}
~~~~~~\XStransition{s_1}{\theta_1}{s'_1} & \bowtie &
\XStransition{s_2}{\theta_2}{s'_2} \\
\hline
\multicolumn{3}{c}{\XStransition{(p,R_1)}{\theta_1~~~~~~~~~}{(p_{\theta_1},R_1
\setminus nodes(\theta_1))}}\\
\end{array}
&
~~~~~~~
&
\begin{array}{l c l}
~~~~~~\XStransition{s_1}{\theta_1}{s'_1} & \bowtie &
\XStransition{s_2}{\theta_2}{s'_2} \\
\hline
\multicolumn{3}{c}{\XStransition{(p,R_2)}{\theta_2~~~~~~~~~}{(p_{\theta_2},R_2\setminus nodes(\theta_2))}}
\end{array}
\end{array}
$
\end{center}
The following example shows the application of the function $div$ to a non-trivial delay-sequence, which contains a combination of sequential and parallel compositions.

\begin{example}
Consider the stochastic Reo connector below. Every indexed $\theta$ is a rate for its respective processing activity, e.g., $\theta_{2}$ is the rate at which the top-left {\sf FIFO1} dispenses data through its sink end; $\theta_3$ is the rate at which the node replicates its coming data, etc. $P_1$ and $P_2$ show up in $\delta_{Macro}$, derived from the Stochastic Reo automaton of this circuit, by two transitions with the delay-sequences of $\lambda_1$ and $\lambda_2$ where: \vspace{-2mm}
\begin{itemize}
 \item from $P_1$: $\lambda_1=((\theta_2; \theta_3)| (\theta_8 ; \theta_9))~;~(\theta_4 | \theta_{10} | \theta_{11})$
 \item from $P_2$: $\lambda_2=(\theta_5 ; \theta_6) ~|~ (\theta_{12} ; \theta_{13})$
\end{itemize}
\vspace{-7mm}
\begin{figure}[!h]
\begin{center}
 \begin{tikzpicture}[font=\tiny]
  \node[point] (a) [label=below right:$\theta_1$] {};
  \node[point] (b) [right of=a,label=below left:$\theta_2$,label=above:$\theta_3$,label=below right:$\theta_4$] {};
  \node[point] (c) [right of=b,label=below left:$\theta_5$] {};
  \node[point] (d) [below of=a,label=above right:$\theta_7$] {};
  \node[point] (e) [right of=d,label=above left:$\theta_8$,label=above right:$\theta_{11}$,label=below:$\theta_9$] {};
  \node[point] (f) [right of=e,label=above left:$\theta_{12}$] {};
  \node[point] (g) [right of=c] {};
  \node[point] (h) [right of=f] {};
  \draw[sync] (c) to node[below] {$\theta_6$} (g);
  \draw[sync] (e) to node[above,xshift=8mm] {$\theta_{13}$} (h);
  \draw[fifo] (a) to (b);
  \draw[syncdrain] (b) to node[left] {$\theta_{10}$} (e);
  \draw[fifo] (d) to (e);	
  \draw[fifo] (b) to (c);
  \draw[fifo] (e) to (f);
  \draw[dashed] (.8,.5) rectangle node[above,yshift=1.2cm] {$P_1$} (2.4,-2.0);
  \draw[dashed] (2.5,.5) rectangle node[above,yshift=1.2cm] {$P_2$} (5,-2.0);
\end{tikzpicture}
 \end{center}
\end{figure}
\vspace{-5mm}
To derive a CTMC, $\lambda_1$ and $\lambda_2$ must be divided into micro-step transitions. We exemplify a few of these divisions. For $\lambda_1$, the division of $(\theta_4 | \theta_{10} | \theta_{11})$ is trivial since it contains only simple parallel composition. This division result is then appended to the division result of $(\theta_2; \theta_3)|(\theta_8 ; \theta_9)$,  which has the same structure as that of $\lambda_2$. Thus, we show below the division result of $\lambda_2$ only.

In the following CTMC fragment, to depict which events have occurred up to a current state, the name of each state shows the delays of all the events that have occurred up to the current state. The delay for a newly occurring event is appended to the existing state name.

\begin{figure}[!h]
\begin{center}
\begin{tikzpicture}[qstate/.style={
		rounded rectangle,
		border, draw=black,
		minimum size=4mm, node distance=2cm},font=\tiny]
 \node[qstate] (a) {$\epsilon ~|~ \epsilon$};
 \node[qstate] (b) [right of=a,yshift=8mm,xshift=-2mm] {$\epsilon  ~|~ \theta_{12}$};
 \node[qstate] (c) [right of=a,yshift=-8mm,xshift=-2mm] {$\theta_5 ~|~ \epsilon$};
 \draw[transition] (a) to (b);
 \draw[transition] (a) to (c);
 \node[qstate] (d) [right of=a,xshift=17mm] {$\theta_5 ~|~ \theta_{12}$};
 \draw[transition] (b) to (d);
 \draw[transition] (c) to (d);
 \node[qstate] (e) [below of=d,yshift=5mm] {$(\theta_5 ; \theta_6) ~|~ \epsilon $};
 \node[qstate] (f) [above of=d,yshift=-5mm] {$\epsilon ~|~(\theta_{12} ; \theta_{13})$};
 \draw[transition] (b) to (f);
 \draw[transition] (c) to (e);
 \node[qstate] (g) [right of=b,xshift=20mm] {$\theta_5~|~ (\theta_{12} ; \theta_{13})$};
 \node[qstate] (h) [right of=c,xshift=20mm] {$(\theta_5 ; \theta_6) ~|~ \theta_{12}$};
 \draw[transition] (f) to (g);
 \draw[transition] (e) to (h);
 \draw[transition] (d) to (g);
 \draw[transition] (d) to (h);
 \node[qstate] (l) [right of=d,xshift=25mm] {$(\theta_5; \theta_6)~|~ (\theta_{12} ; \theta_{13})$};
 \draw[transition] (g) to (l);
 \draw[transition] (h) to (l);
\end{tikzpicture}
\end{center}
% \caption{Division result of the transition consuming data from the buffers in Figure~\ref{fig:example}}
\label{fig:exampleSeq}
\end{figure}
\vspace{-5mm}

This example shows that when a delay-sequence $\lambda$ is generated by parallel composition, the events in one of the sub-delay-sequences of $\lambda$ occur independently of the events in other sub-delay-sequences. Still they keep their occurrence order in the sub-delay-sequence that they belong to. $\blacksquare$
\end{example}

The division into micro-step transitions ensures that each transition has a single 3-tuple in its label. Thus, the micro-step transitions can be extracted as:%[Processing of single data flow]
\begin{center}
$\delta_{Proc}=\{\XStransition{(p,R)}{v(\theta)~}{(p',R')}
\mid
\XStransition{(p,R)}{\theta}{(p',R')} \in div(t) ~\textrm{for all } t \in \delta_{Macro}\}
$
\end{center}

Synchronized data-flows in a Reo automaton are
considered atomic, hence other events cannot interfere with them. However, splitting these data-flows allows non-interfering events to interleave with their micro-steps, disregarding the strict sense of their atomicity. For example, a certain boundary node unrelated to a group of synchronized data-flows can
accept a data item between any two micro-steps. Since we want to allow such interleaving, we must explicitly add such
data-arrivals. For a Stochastic Reo automaton $(\mathcal A,\mathbf r,\mathbf t)$ with
$\mathcal A=(\Sigma,Q,\delta_{\mathcal A})$ and a set of micro-step states $S_M$,
its full set of data-arrival transitions, including its data-arrivals, is defined as:
\begin{center}
$\delta_{Arr}=\delta'_{Arr} \cup
\{\XStransition{(p,R)}{\mathbf r(d)}{(p,R\cup \{d\})}
\mid
(p,R),(p,R\cup \{d\}) \in S_M,~d \in \Sigma,~d \notin R\}
$
\end{center}

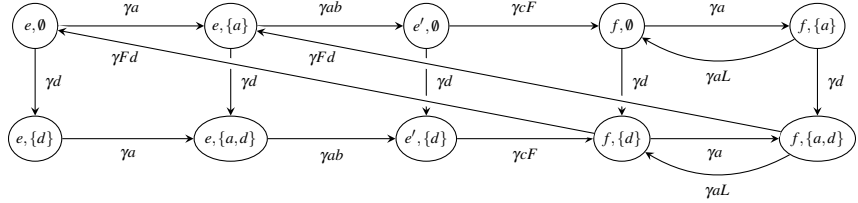
\begin{figure}[!h]
\begin{center}
\begin{tikzpicture}[font=\tiny,transition/.style={->,>=stealth},
 					state/.style={ellipse,inner sep=0pt,draw,minimum size=6mm,node
 					distance=1.8cm}]
\node[state] (A) {$e,\emptyset$};
\node[state] (B) [right of=A,xshift=.8cm] {$e,\{a\}$};
\node[state] (C) [right of=B,xshift=.8cm] {$e',\emptyset$};
\node[state] (D) [right of=C,xshift=.8cm] {$f,\emptyset$};
\node[state] (E) [right of=D,xshift=.8cm] {$f,\{a\}$};
\node[state] (F) [below of=A,yshift=.3cm] {$e,\{d\}$};
\node[state] (G) [below of=B,yshift=.3cm] {$e,\{a,d\}$};
\node[state] (H) [below of=C,yshift=.3cm] {$e',\{d\}$};
\node[state] (I) [below of=D,yshift=.3cm] {$f,\{d\}$};
\node[state] (J) [below of=E,yshift=.3cm] {$f,\{a,d\}$};
\draw[transition] (A) to node[above] {$\gamma a$} (B);
\draw[transition] (B) to node[above] {$\gamma ab$} (C);
\draw[transition] (C) to node[above] {$\gamma cF$} (D);
\draw[transition] (D) to node[above] {$\gamma a$} (E);
\draw[transition] (F) to node[below] {$\gamma a$} (G);
\draw[transition] (G) to node[below] {$\gamma ab$} (H);
\draw[transition] (H) to node[below] {$\gamma cF$} (I);
\draw[transition] (I) to node[below] {$\gamma a$} (J);
\draw[transition] (A) to node[right] {$\gamma d$} (F);
\draw[transition] (B) to node[right] {$\gamma d$} (G);
\draw[transition] (C) to node[right] {$\gamma d$} (H);
\draw[transition] (D) to node[right] {$\gamma d$} (I);
\draw[transition] (E) to node[right] {$\gamma d$} (J);
\draw[-,line width=4pt,draw=white] (I) to (A);
\draw[transition] (I) to node[very near end,yshift=-.2cm] {$\gamma Fd$} (A);
\draw[-,line width=4pt,draw=white] (J) to (B);
\draw[transition] (J) to node[very near end,yshift=-.2cm] {$\gamma Fd$} (B);
\draw[transition] (J) [bend left] to node[below] {$\gamma aL$} (I);
\draw[transition] (E) [bend left] to node[below] {$\gamma aL$} (D);
\end{tikzpicture}
\end{center}
\vspace{-7mm}
\caption{Derived CTMC of {\sf LossyFIFO1}}
\label{fig:ctmc}
\end{figure}

\begin{figure}[!h]
\begin{center}
 \includegraphics[scale=0.3]{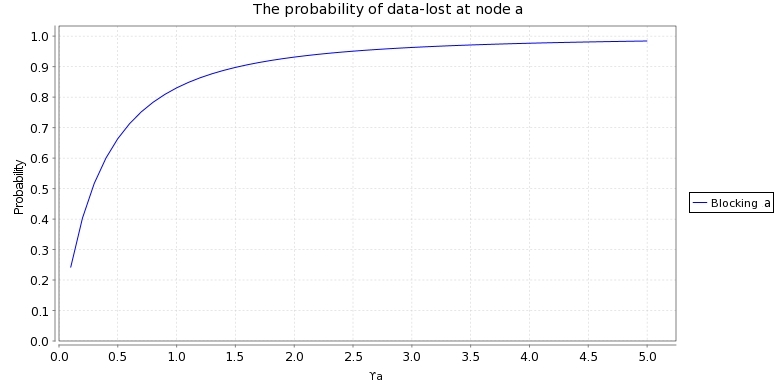}
\end{center}
\vspace{-4mm}
\caption{Probability of data lost at node \emph{a}}
\label{fig:dataLost}
\end{figure}

The derived CTMC model can be used for stochastic analysis. For instance, Figure~\ref{fig:dataLost} is obtained from PRISM~\cite{PRISM} using the CTMC model~(see Figure~\ref{fig:ctmc}) derived from the Stochastic Reo circuit of the {\sf LossyFIFO1} example in Figure~\ref{fig:stochasticlossyfifo}. Figure~\ref{fig:dataLost} shows how the probability of data loss varies as the arrival rate at node $a$ increases.

% Figure \ref{fig:ctmc} shows the derived CTMC for the {\sf LossyFIFO1} circuit in Figure \ref{fig:stochasticlossyfifo}.

%The sets of states and transitions of the CTMC model are:
% $\begin{array}{ll}
% S_A= & \{(e,\emptyset),(e,\{d\}),(e,\{a\}),(e,\{a,d\}),(f,\emptyset),(f,\{d\}),(f,\{a\}),(f,\{a,d\})\}\\
% S_M= & \{(e',\emptyset),(e'\{d\})\} \\
% \delta'_{Arr} = & \{\} \\
% \delta'_{Arr}
% \end{array}$
\vspace{-3mm}
\vspace{-5mm}

\section{Related work} \label{s.RelatedWork}
The research in formal specification of systems with quantitative
aspects encompasses a variety of developments such as Stochastic Process
Algebras (SPAs) \cite{CT02}, Stochastic Automata Networks (SANs)
\cite{FPS98,SAP95}, and Stochastic Petri nets (SPNs)
\cite{HMRT01,STP96}. %Measure Specification Language (MSL) \cite{AB06} ,
SPA is a model for both qualitative and quantitative specification and analysis with a compositional and hierarchical framework. It has algebraic laws (the so-called static laws) and expansion laws which express parallel compositions in terms of SPA operators. In SPA the interpretation of the parallel composition is a vexed one because it allows various interpretations such as Performance Evaluation Process Algebra (PEPA) \cite{H96}, and Extended Markovian Process Algebra (EMPA) \cite{BG96,BG98}. SPA describes `\textit{how}' each process behaves, while (Stochastic) Reo directly describes `\textit{what}' communication protocols connect and coordinate the processes in a system, in terms of primitive channels and their composition. Therefore, (Stochastic) Reo explicitly models the pure coordination and communication protocols including the impact of real communication networks on software systems and their interactions. Compared to SPA, our approach more naturally leads to a formulation using queueing models.

SPN is widely used for modelling concurrency, synchronization, and precedence, and is conducive to both top-down and bottom-up modelling. Stochastic Reo shares the same properties with SPN and natively supports composition of synchrony and exclusion together with asynchrony. The topology of connectors in (Stochastic) Reo is inherently dynamic, and it accommodates mobility~\cite{KMLA10}. Moreover, (Stochastic) Reo supports a liberal notion of channels and is more general than data-flow models and Petri nets, which can be viewed as specialized channel-based models that incorporate certain specific primitive coordination constructs.

SAN consists of a couple of stochastic automata that act independently. Thus, the state of SAN at time $t$ is expressed by the states of each automaton at time $t$. The concept of a collection of individual automata helps modeling distributed and parallel systems more easily. The interactions in SAN are rather limited to patterns like synchronizing events or operating at different rates. Compared with the SAN approach, the expressiveness of (Stochastic) Reo makes it possible to model different interaction patterns involving both asynchronous and synchronous communications.

Continuous-Time Constraint Automata~(CCA)~\cite{CCA06} are another stochastic extension of CA that support reasoning about QoS aspects such as expected response times. CCA are close to Interactive Markov Chains~(IMCs)~\cite{IMC02}, that is, they have two types of transitions called \emph{interactive transitions} and \emph{Markovian transitions} for, respectively, the immediate interaction with the environment and internal stochastic behaviors. The stochastic extension in CCA focuses on internal behavior of a connector, but does not take into account the arrivals of I/O requests at the ends of a connector as a stochastic process, which is required for reasoning about the end-to-end QoS of a system.

\section{Conclusions and Future work}

We introduced Stochastic Reo automata by extending Reo automata with
functions that assign stochastic values of arrival rates and processing delay
rates to boundary nodes and channels in Stochastic Reo. This model is very compact compared to
the existing models, e.g., in~\cite{QIA09}. Various formal properties of our model are obtained, reusing the formal justifications of the various properties of Reo automata \cite{BCS09}, such as compositionality.

The technical core in this paper shows the complexity of the original problem whence it stems from: derivation of stochastic models for formal analysis of end-to-end QoS properties of systems composed of services/components supplied by disparate providers, in their user environments. This complexity highlights the gross inadequacy of informal, or one-off techniques and emphasizes the importance of formal approaches and sound models that can serve as the basis for automated tools. %Stochastic Reo automata presented in this paper form the basis of an integrated set of tools, Eclipse Coordination Tools~(ECT)~\cite{ECT}, that we have implemented for this purpose. %Using ECT, we can model service/component connectors as Stochastic Reo circuits, compositionally derive their semantics as Stochastic Reo automata, and translate the results into input for third-party tools for stochastic analysis, such as PRISM, MATLAB, or Maple.

Stochastic Reo does not impose any restriction on the distribution of its annotated rates such as the rates for data-arrivals at channel ends or data-flows through channels. However, for translation of stochastic Reo to a homogeneous CTMC model, we considered only the exponential distributions for the rates. As future work, we also want to consider non-exponential distributions by considering phase-type distributions \cite{O90} or using Semi-Markov Processes \cite{YS04} as target
models of our translation. A simulation engine~\cite{oscarSimulation10}, already integrated into our toolset, Eclipse Coordination Tools~(ECT)~\cite{ECT} environment, supports a wide variety of more general distributions for Stochastic Reo. In our future work, we will consider rewards of a system along with its stochastic behavior as well. Our translation result will thus become a CTMC model with reward information on its transitions and states, and can be fed into a stochastic analysis tool. We will also implement a tool for our translation approach via Stochastic Reo automata. It will be an extension of the existing tools in ECT, for instance, by implementing the synchronization in Stochastic Reo automata. Furthermore, as a case study, we are currently applying our method to an industrial application, the \emph{ASK} system \cite{ASK}, by modeling the system with Stochastic Reo and analyzing the CTMC derived from its resulting Stochastic Reo automaton model~\cite{Example}.

\section{Acknowledgments}
We thank the anonymous reviewers and Michel Reniers for their valuable comments and suggestions.

\bibliographystyle{eptcs} % or whatever you prefer

\begin{thebibliography}{1}
\bibitem{BSAR06}
Christel Baier, Marjan Sirjani, Farhad Arbab \& Jan J. M. M. Rutten (2006):
 \newblock \emph{Modeling {C}omponent {C}onnectors in {R}eo by {C}onstraint {A}utomata}.
 \newblock {\sl Sci. Comput. Program.} 61, pp. 75--113 .

\bibitem{Arbab04}
Farhad Arbab (2004):
  \newblock \emph{Reo: {A} {C}hannel-based {C}oordination {M}odel for {C}omponent {C}omposition}.
  \newblock {\sl MSCS} 14, pp. 329--366.

\bibitem{BCS09}
Marcello Bonsangue, Dave Clarke, \& Alexandra Silva (2009):
 \newblock \emph{Automata for {C}ontext-{D}ependent {C}onnectors}.
 \newblock {Coordination {M}odels and {L}anguages}.
 \newblock {\sl LNCS} 5521. pp. 184--203.
 \newblock {Springer Verlag}

\bibitem{QIA09}
Farhad Arbab, Tom Chothia, Rob van de Mei, Sun Meng, Young-Joo Moon \& Chr\'{e}tien Verhoef (2009):
 \newblock {\em {F}rom {C}oordination to {S}tochastic {M}odels of {QoS}}.
 \newblock {Coordination {M}odels and {L}anguages}.
 \newblock {\sl LNCS} 5521. pp. 268--287.
 \newblock {Springer Verlag}.

\bibitem{CCA05}
Dave Clarke, David Costa \& Farhad Arbab (2007):
  \newblock {\em Connector Colouring I: Synchronisation and Context Dependency}.
  \newblock {\sl Sci. Comput. Program.} 66. pp. 205--225.

\bibitem{CT02}
Mariacarla Calzarossa \& Salvatore Tucci (2002):
  \newblock {\em Performance Evaluation of Complex Systems: Techniques and Tools, Performance 2002, Tutorial Lectures}.
  \newblock {Performance}.
  \newblock {\sl LNCS} 2459.
  \newblock {Springer}.

\bibitem{FPS98}
Paulo Fernandes, Brigitte Plateau \& William J. Stewart (1998):
  \newblock {\em Efficient Descriptor-Vector Multiplications in Stochastic Automata Networks}.
  \newblock {\sl J. ACM} 45. pp. 381--414.

\bibitem{SAP95}
William Stewart, Karim Atif \& Brigitte Plateau (1995):
  \newblock {\em The Numerical Solution of Stochastic Automata Networks}.
  \newblock {\sl EOR} 86. pp. 503--525.

\bibitem{HMRT01}
Boudewijn R. Haverkort, Raymond Marie, Gerardo Rubino \& Kishor Shridharbhai Trivedi (2001):
  \newblock {\em Performability Modelling: Techniques and Tools}.
  \newblock {\sl Wiley}.

\bibitem{STP96}
Robin A. Sahner, Kishor S. Trivedi \& Antonio Puliafito (1996):
  \newblock {\em Performance and reliability analysis of computer systems: an example-based approach using the SHARPE software package}.
  \newblock {\sl Kluwer Academic Publishers}.

\bibitem{H96}
J. Hillston (1996):
  \newblock {\em A Compositional Approach to Performance Modelling}.
  \newblock {\sl Cambridge University Press}.

\bibitem{BG96}
Marco Bernardo \& Roberto Gorrieri (1996):
  \newblock {\em Extended Markovian Process Algebra}.
  \newblock {CONCUR}.
  \newblock {\sl LNCS} 1119. pp. 315--330.
  \newblock {Springer}.


\bibitem{BG98}
Marco Bernardo \& Roberto Gorrieri (1998):
  \newblock {\em A Tutorial on EMPA: A Theory of Concurrent Processes with Nondeterminism, Priorities, Probabilities and Time}.
  \newblock {\sl Theor. Comput. Sci.} 202. pp1--54.

\bibitem{O90}
Colm A. O'Cinneide (1990):
  \newblock {\em Characterization of phase-type distributions}.
  \newblock {Stochastic Models} 6. pp. 1--57.

\bibitem{YS04}
H{\aa}kan L. S. Younes \& Reid G. Simmons (2004):
  \newblock {\em Solving Generalized Semi-Markov Decision Processes Using Continuous Phase-Type Distributions}.
  \newblock {AAAI}. pp. 742--748.

\bibitem{ASK}
Andries Stam (2008):
  \newblock {\em The ASK System and the Challenge of Distributed Knowledge Discovery}.
  \newblock {ISoLA}.
  \newblock {\sl Communications in Computer and Information Science} 17. pp. 663--668.
  \newblock {Springer}.

\bibitem{PRISM}
Probabilistic model checker:
 \newblock {http://www.prismmodelchecker.org/}.

\bibitem{ECT}
Eclipse {C}oordination {T}ools:
  \newblock {http://reo.project.cwi.nl}.

\bibitem{IMC02}
Holger Hermanns (2002):
  \newblock {\em Interactive Markov Chains: The Quest for Quantified Quality}.
  \newblock {\sl Lecture Notes in Computer Science} 2428.
  \newblock {Springer}.

\bibitem{CCA06}
Christel Baier \& Verena Wolf (2006):
  \newblock {\em Stochastic Reasoning About Channel-Based Component Connectors}.
  \newblock {COORDINATION}.
  \newblock {\sl LNCS} 4038. pp. 1--15.
  \newblock {Springer}.

\bibitem{oscarSimulation10}
Oscar Kanters (2010):
 \newblock {\em QoS analysis by simulation in Reo}.
 \newblock {Vrije Universiteit}
 \newblock {Amsterdam, The Netherlands}.

\bibitem{KMLA10}
Christian Krause, Ziyan Maraikar, Alexander Lazovik \& Farhad Arbab (2010):
  \newblock {\em Modeling Dynamic Reconfigurations in {Reo} using High-Level Replacement Systems}.
  \newblock {Science of Computer Programming}.
  \newblock {\sl Elsevier}.
  \newblock {To appear}.

\bibitem{Example}
Case study in ASK system:
  \newblock{http://reo.project.cwi.nl/cgi-bin/trac.cgi/reo/wiki/CaseStudies/StochasticReoASK}

\end{thebibliography}

\end{document}